\documentclass[twoside,a4paper,11pt]{proceedings}
\usepackage{graphicx}
\usepackage{hyperref}
\usepackage{movie15}
\usepackage{natbib}
\begin{document}
\pagenumbering{arabic}
\pagestyle{myheadings}
\thispagestyle{empty}
\vspace*{-1cm}
{\flushleft\includegraphics[width=8cm,viewport=0 -30 200 -20]{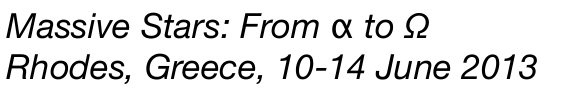}}
\vspace*{0.2cm}
\begin{flushleft}
{\bf {\LARGE
Galactic Evolved Massive Stars Discovered by Their Infrared Emission
}\\
\vspace*{1cm}
A. P. Marston$^1$,
J. Mauerhan$^2$,
S. Van Dyk$^3$,
M. Cohen$^4$
and
P. Morris$^5$
%
}\\
\vspace*{0.5cm}
%
$^{1}$
ESA/ESAC, PO Box 78, 28691 Villanueva de la Ca\~nada, Madrid, Spain \\
$^{2}$
Steward Observatory, U. Arizona, Tucson, AZ 85721-0065, USA \\
$^{3}$
Spitzer Science Center/Caltech, 220-6, Pasadena, CA 91125, USA \\
$^{4}$
Monterey Institute for Research in Astronomy, 200 8th Street, Marina, CA 93933, USA \\
$^{5}$
NASA Herschel Science Center/Caltech, 100-22, Pasadena, CA 91125, USA
%
\end{flushleft}
\markboth{
Galactic Evolved Stars from Their Infrared Emission
}{
Marston et al.
}
\thispagestyle{empty}
\vspace*{0.4cm}
\begin{minipage}[l]{0.09\textwidth}
\ 
\end{minipage}
\begin{minipage}[r]{0.9\textwidth}
\vspace{1cm}
\section*{Abstract}{\small
Determining the Galactic distribution and numbers of massive stars, such as Wolf-Rayet stars (WRs), is hampered by intervening Galactic or local circumstellar dust obscuration. In order to probe such regions of the Galaxy we can use infrared observations, which provide a means for finding such hidden populations through the dust.
The availability of both 2MASS and Spitzer/GLIMPSE large-scale survey data provides infrared colours from 1.25 to 8$\mu$m for a large fraction of the inner Galactic plane. In 2005 we initiated a pilot study of the combined set of infrared colours for two GLIMPSE fields and showed that WRs typically occupy a sparsely populated region of the colour space.
We followed up 42 of our WR candidates spectroscopically in the near-infrared, and with limited additional observations of some of these candidates in the optical. 
Six new WRs, four late-type WN and two late-type WC stars, were discovered as a result.  Of the remaining $\sim$86\% of the sample, five appear to be O-type stars. 21 stars are likely of type Be, and 10 stars appear to be of late-type, or possibly young stellar objects, which have ``contaminated'' the infrared color space.
The survey is generally unbiased towards clusters or field stars, and the new WRs found are in both the field and in and around the RCW 49 region (including cluster Westerlund 2). In this work, and in our other recent work, we show that the infrared broad-band colours to be the most efficient means of identifying (particularly, dust-obscured) candidate massive stars, notably WRs.

\vspace{10mm}
\normalsize}
\end{minipage}
\section{Introduction}
Wolf-Rayet (WR) stars represent the later evolutionary stages of massive stars (single or in binaries) with initial main sequence masses larger than $20 M_{\odot}$. The exact evolutionary paths for such high mass stars are not well established and there are several possibilities depending on various factors such as initial mass, rotation rate and magnetic field \citep[See][]{langer12} for a recent overview of the situation. Clearly, better statistical information covering a range of stellar parameters can help in the understanding of likely evolutionary scenarios for very massive stars. In the last 10 years or so the number of known WR stars has more than doubled \citep{shara09,shara2012,hadfield07,mauerhan11}, with more than 430 galactic WR stars now known. This article reports on early, previously unreported, work in identifying WR stars in the galaxy. 
\section{Candidate Galactic WR Stars}
\begin{figure}[!ht]
\begin{center}
      \resizebox{0.4\hsize}{!}{
      \includegraphics*{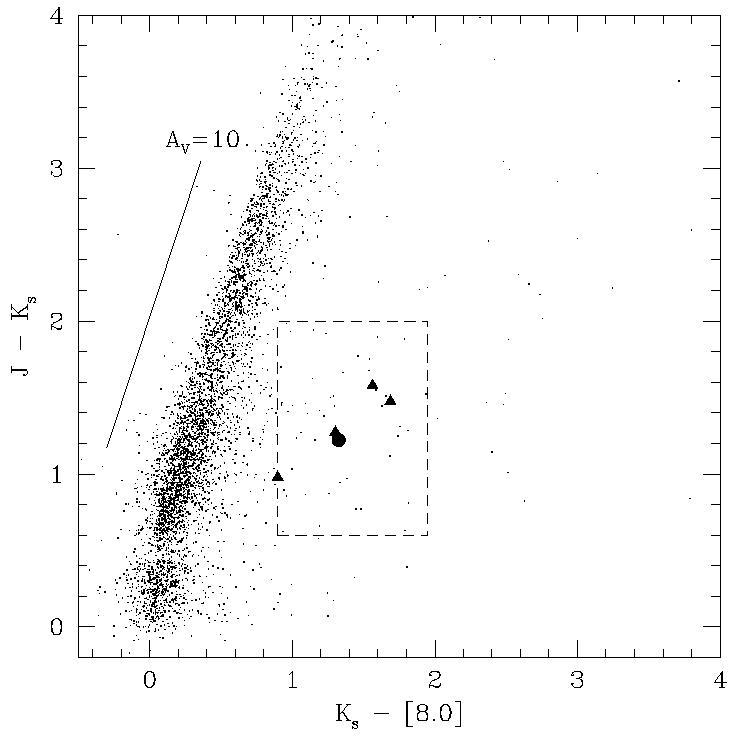}
}
\end{center}
\caption{Colour-colour diagram of 2$^{\deg}$ by 2$^{\deg}$ segment of the GLIMPSE catalogue, centred on the H II region RCW 49, cross-matched (with a 1$^{\prime\prime}$ match 
radius) with the highest-reliability 2MASS point sources ($\sim$37,500 matches), shown as points in the diagram. Also shown are the previously known WRs in RCW 49 (WNs = triangles; WCs =  circles). 
}
\label{fig:ccdiag}
\end{figure}

The strong winds and significant mass loss of WR stars lead to a power-law spectrum excess from the wind with spectral index of 2.7 to 3.2 which is in addition to the photospheric emission \citep{morris93}. This becomes particularly notable in the near- to mid-infrared \citep[See for more details][]{mauerhan11}.  The Spitzer/GLIMPSE survey of the galactic plane at galactic latitudes $-1\deg < b < +1\deg$ provides infrared fluxes for tens of millions of sources in the wavelength range 3.8$\mu$m to 8$\mu$m {benjamin03}. The 2MASS survey covers the whole sky at the near-infrared JHK$_s$ wavebands. Combining GLIMPSE and 2MASS colours allows a colour-colour diagram of sources where WR stars are in a clearly defined colour space containing a relatively low percentage of objects (Figure~\ref{fig:ccdiag}).  The reddening vector shows that WR stars remain in a clearly distinguishable region of the colour-colour diagram.

\begin{figure}[!ht]
\begin{center}
    {
  \centering 
  \leavevmode
  \includegraphics[width=.45\textwidth]{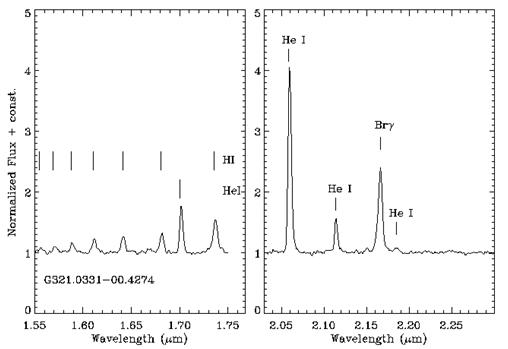}%
  \hfil
  \includegraphics[width=.45\textwidth]{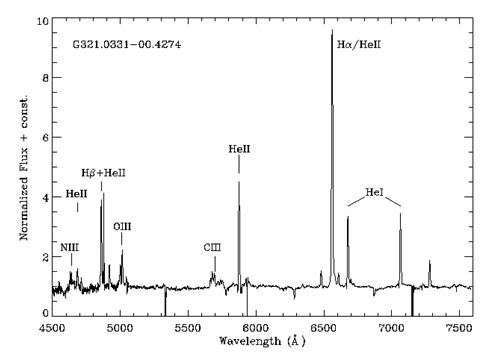}%
}
\end{center}
\caption{{\it Left}: The normalized H and K SOFI/NTT spectra for the new WN10 star discovered in this program, G321.0331-00.4274 {\it Right:} Normalized optical spectrum of the same star taken at the CTIO 4m Blanco telescope and RC Spec.
}\label{fig:wn10}
\end{figure}

In the initial pilot project reported here, candidates were chosen from a 2$\deg$ region around RCW~49 and a relatively random region of the galaxy around a galactic latitude of $318\deg < l < 320\deg$. These were followed up with spectra taken with the SOFI instrument on board the New Technology Telescope (NTT) on La Silla, Chile, effectively covering the 1.8 to 2.5 $\mu$m spectral range in June 2005. This covers a number diagnostic H, He and C line wavelengths for identifying WN and WC subtypes of WR stars. Since only spectral line diagnostics were needed, flux calibration was not required and weather conditions were variable. Further optical spectroscopy followup was performed on the CTIO 4m Blanco telescope using the RC spectrometer in March 2006.

\section{Spectral results}
Our spectral follow-up revealed a $\sim15$\% success rate for finding WR stars. A total of 6 new WR stars were found from 42 candidate stars viewed. Four of these new WR stars were found away from the Westerlund 2 cluster/RCW49 field while two further WR stars were found in the RCW49 region outer regions.  These include 2 WC stars and 4 WN stars, including the new WN10 star G321.0331-00.4274 (see Figure~\ref{fig:wn10}).

In total, 80\% of stars showed emission lines. Most of these are Be (or sometimes O) stars. Examples are shown in Fig.A1 of \citet{mauerhan11} where the H and K band spectra are dominated by HI emission lines.

\section{Conclusions}

Later work has improved the colour selection criterion of the method used here \citet{mauerhan11} and a significant number of candidates are still to be looked at. \citet{rlopes11a,rlopes11b,rlopes11c} have presented the discovery SOFI infrared spectra of 5 of the 6 WR stars discovered in this study using SOFI archival data. These are referred to as WR20aa, WR20c, WR60aa, WR67a, and WR67b. Identifying WR star candidates via their infrared colours has since been improved \citep{hadfield07,mauerhan09,mauerhan11}.

One of the curious aspects of this work -- as has also been seen in the further studies of \citet{hadfield07} and \citet{mauerhan11} -- is the fact that as many newly discovered WR stars are found well away from dense star clusters as associated with them. Since their lives are so short many of these objects may be formed {\it in situ} or very near by. Alternately, there is a large fraction of WR stars forming a population of fast moving, runaway stars in the galaxy. Something that may well be resolved by upcoming GAIA observations.

\small  
%
%

{}
\end{document}